\documentclass[prb,citeautoscript,twocolumn,showpacs,superscriptaddress]{revtex4-2} 
\usepackage{graphicx}
\usepackage{amsmath}
\usepackage{color}
\usepackage{float}
\usepackage{placeins}
\usepackage[retainorgcmds]{IEEEtrantools}
\usepackage{natbib}
\usepackage{mwe}
\usepackage[caption=false]{subfig} 
\usepackage{ragged2e} 
\DeclareCaptionJustification{justified}{\justifying}
\usepackage{adjustbox}

\begin{document}

\newcommand{\rt}{R$_2$T$_2$O$_7$}
\newcommand{\lahf}{La$_2$Hf$_2$O$_7$}
\newcommand{\cesn}{Ce$_2$Sn$_2$O$_7$}
\newcommand{\cezr}{Ce$_2$Zr$_2$O$_7$}
\newcommand{\cehf}{Ce$_2$Hf$_2$O$_{7}$}
\newcommand{\przr}{Pr$_2$Zr$_2$O$_7$}
\newcommand{\prsn}{Pr$_2$Sn$_2$O$_7$}
\newcommand{\prhf}{Pr$_2$Hf$_2$O$_7$}
\newcommand{\ndzr}{Nd$_2$Zr$_2$O$_7$}
\newcommand{\hoti}{Ho$_2$Ti$_2$O$_7$}
\newcommand{\dyti}{Dy$_2$Ti$_2$O$_7$}
\newcommand{\tbti}{Tb$_2$Ti$_2$O$_7$}
\newcommand{\tbsn}{Tb$_2$Sn$_2$O$_7$}
\newcommand{\ce}{Ce$^{3+}$}
\newcommand{\cebis}{Ce$^{4+}$}
\newcommand{\pr}{Pr$^{3+}$}
\newcommand{\yb}{Yb$^{3+}$}
\newcommand{\degC}{$^\circ$C}

\title{Crystal-field states and defect levels in candidate quantum spin ice \cehf}
\author{Victor Por\'{e}e}
\email[]{victor.poree@psi.ch}
\affiliation{Laboratory for Neutron Scattering and Imaging, Paul Scherrer Institut, 5232 Villigen, Switzerland}
\author{Elsa Lhotel}
\affiliation{Institut N\'eel, CNRS, Universit\'e Grenoble Alpes, 38042 Grenoble, France}
\author{Sylvain Petit}
\affiliation{LLB, CEA, CNRS, Universit\'{e} Paris-Saclay, CEA Saclay, 91191 Gif-sur-Yvette, France}
\author{Aleksandra Krajewska}
\affiliation{ISIS Neutron and Muon Source, Rutherford Appleton Laboratory, Chilton, Didcot, OX11 OQX, United Kingdom}
\author{Pascal Puphal}
\affiliation{Laboratory for Multiscale Materials Experiments, Paul Scherrer Institut, Villigen CH-5232, Switzerland}
\affiliation{Max Planck Institute for Solid State Research, Heisenbergstrasse 1, 70569 Stuttgart, Germany}
\author{Adam H. Clark}
\affiliation{Swiss Light Source, Paul Scherrer Institut, 5232 Villigen, Switzerland}
\author{Vladimir Pomjakushin}
\affiliation{Laboratory for Neutron Scattering and Imaging, Paul Scherrer Institut, 5232 Villigen, Switzerland}
\author{Helen C. Walker}
\affiliation{ISIS Neutron and Muon Source, Rutherford Appleton Laboratory, Chilton, Didcot, OX11 OQX, United Kingdom}
\author{Nicolas Gauthier}
\affiliation{Stanford Institute for Materials and Energy Science, SLAC National Accelerator Laboratory and Stanford University, Menlo Park, California 94025, USA}
\author{Dariusz J. Gawryluk}
\affiliation{Laboratory for Multiscale Materials Experiments, Paul Scherrer Institut, Villigen CH-5232, Switzerland}
\author{Romain Sibille}
\email[]{romain.sibille@psi.ch}
\affiliation{Laboratory for Neutron Scattering and Imaging, Paul Scherrer Institut, 5232 Villigen, Switzerland}

\begin{abstract}
We report the synthesis of powder and single-crystal samples of the cerium pyrohafnate and their characterization using neutron diffraction, thermogravimetry and X-ray absorption spectroscopy. We evaluate the amount of non-magnetic \cebis~defects and use this result to interpret the spectrum of crystal-electric field transitions observed using inelastic neutron scattering. The analysis of these single-ion transitions indicates the dipole-octupole nature of the ground state doublet and a significant degree of spin-lattice coupling. The single-ion properties calculated from the crystal-electric field parameters obtained spectroscopically are in good agreement with bulk magnetic susceptibility data down to about 1~K. Below this temperature, the behavior of the magnetic susceptibility indicates a correlated regime without showing any sign of magnetic long-range order or freezing down to 0.08~K. We conclude that \cehf~is another candidate to investigate exotic correlated states of quantum matter such as the octupolar quantum spin ice recently argued to exist in the isostructural compounds \cesn~and \cezr.
\end{abstract}

\maketitle
\section{introduction}
Exotic magnetic behaviors often stem from competing interactions~\cite{Lhuillier:BEVn7Mgc}. This was studied in geometrically frustrated systems, such as triangular~\cite{AFM_Triangle}, kagome~\cite{MENDELS_2016}, hyperkagome~\cite{Deen_2010,Paddison179} and pyrochlore~\cite{PyroStr1} lattices. On the pyrochlore lattice, magnetic moments reside on the nodes of a network of corner-sharing tetrahedra. Under certain conditions, such systems can enter a spin ice state, where two spins point towards the tetrahedron center and the two others away from it. A few rare-earth pyrochlore oxides, such as \hoti~\cite{PRL_HTO,Fennell415,Fennell_PRB} or \dyti~\cite{Morris411,Fennell_PRB}, were found to host real life realizations of spin ices.

The epitome of frustrated systems is incarnated by the quantum spin liquid (QSL) states~\cite{Balents:2010jx}. Owing to quantum fluctuations, QSLs are expected to evade ordering or freezing of the magnetic moments down to $T$~=~0~K. Instead, QSLs are long-range quantum entangled and have deconfined excitations. Such a phase on the pyrochlore lattice is called a quantum spin ice (QSI)~\cite{Gingras:2014ip} and results from the coherent superposition of `2-in-2-out' spin ice configurations. Theory has identified several possible ways to stabilize a QSI. The existence of low-lying crystal-electric field states was first argued to be possible ingredients~\cite{Molavian2007} in the context of understanding the absence of magnetic order in \tbti. Multipolar interactions provide another possible route, as was proposed for praseodymium compounds~\cite{Onoda2010}. In this case, interactions between electric quadrupoles play the role of a transverse exchange allowing fluctuations among spin ice states of Ising magnetic dipoles. A number of experimental works~\cite{Zhou:2008cz,PZO_Sylvain,Kimura:2013gj,PHO_PRB,PHO_NatPhy} have therefore concentrated on studying Pr$^{3+}$-based pyrochlore materials. Finally, it has also been proposed that non-magnetic disorder could promote a QSI via the introduction of transverse fields in the Hamiltonian~\cite{Disorder_Savary,Disorder_Benton,Wen_PZO_PRL}.

While multipoles have been proposed as a source of transverse terms in models dominated by couplings among magnetic dipoles~\cite{Onoda2010}, correlated phases of higher-rank multipoles have also been conjectured for pyrochlores where the trivalent rare earth has a dipole-octupole ground-state doublet~\cite{DO-PRL,DO-PRB}. Such single-ion ground states are notably possible in pyrochlore materials based on cerium, samarium or neodymuim~\cite{DO-PRL,DO-PRB,Benton_DO}. In particular, both \cesn~\cite{CSO_PRL,CSO_NatPhy} and \cezr~\cite{CZO_CA,CZO_US} are identified as dipole-octupole pyrochlores. \cesn~was proposed to form a peculiar $U(1)$ quantum spin liquid state where couplings between the octupolar components of the pseudo-spins dominate while couplings between the dipole components play the role of transverse fluctuations, resulting in a QSI based on a manifold of octupole ice configurations~\cite{DO-PRL,DO-PRB}. Ultimately this was demonstrated by the observation of a specific signature in neutron scattering - the rise of a weak diffuse scattering signal occurring at high momentum transfer due to the peculiar magnetization density of octupoles. The octupolar moment develops at the expense of the dipole moment upon cooling inside this phase, due to the effect of the dominant octupolar correlations on the wavefunctions. The difficult synthesis of single crystals of \cesn, though remarkable progress was made recently~\cite{CSO_crystal}, however limits more detailed studies. Instead, large single-crystals of \cezr~were successfully grown and investigated using macroscopic probes as well as neutron scattering techniques~\cite{CZO_CA,CZO_US}. The magnetic properties of \cezr~share similarities with those of \cesn$ $, particularly the continua of low-energy magnetic excitations identified in both compounds. Recent works~\cite{PyrochloreU(1)GangChen,CZO_US_2,CZO_CA_2}, modelling the macroscopic properties and the momentum-resolved inelastic neutron scattering of \cezr, seem to confirm that the octupolar quantum spin ice first unveiled in \cesn~\cite{CSO_NatPhy} is common among cerium pyrochlores. Overall, there is growing evidence that \ce~pyrochlores are concrete realisations of a 3D QSL~\cite{CZO_US,CSO_NatPhy,CZO_US_2,CZO_CA_2}, which calls for further efforts in investigating such materials. 

\cehf~is another pyrochlore compound where trivalent cerium ions can be stabilized. Here we report the preparation and characterization of Ce$_2$Hf$_2$O$_{7+\delta}$, focusing on the level of residual tetravalent cerium $\delta$ achieved in our samples. We present a detailed study of the single-ion properties via crystal electric field analysis and compare our results with magnetic measurements performed on powder samples. Although further work is required to reduce $\delta$ well below 0.1 in single crystals, their availability opens the door to measuring both the high-momentum diffuse neutron scattering specific of the octupolar-QSI and the associated low-energy excitations in a momentum-resolved manner. Our magnetic susceptibility data additionally show that the behavior observed in \cesn, namely the progressive reduction of dipole magnetic moment upon cooling below 1 K, is also observed in \cehf, revealing a possible indication of an octupolar-QSI ground state.

\section{Methods}

\begin{figure}[!hb]
\includegraphics[width=\linewidth]{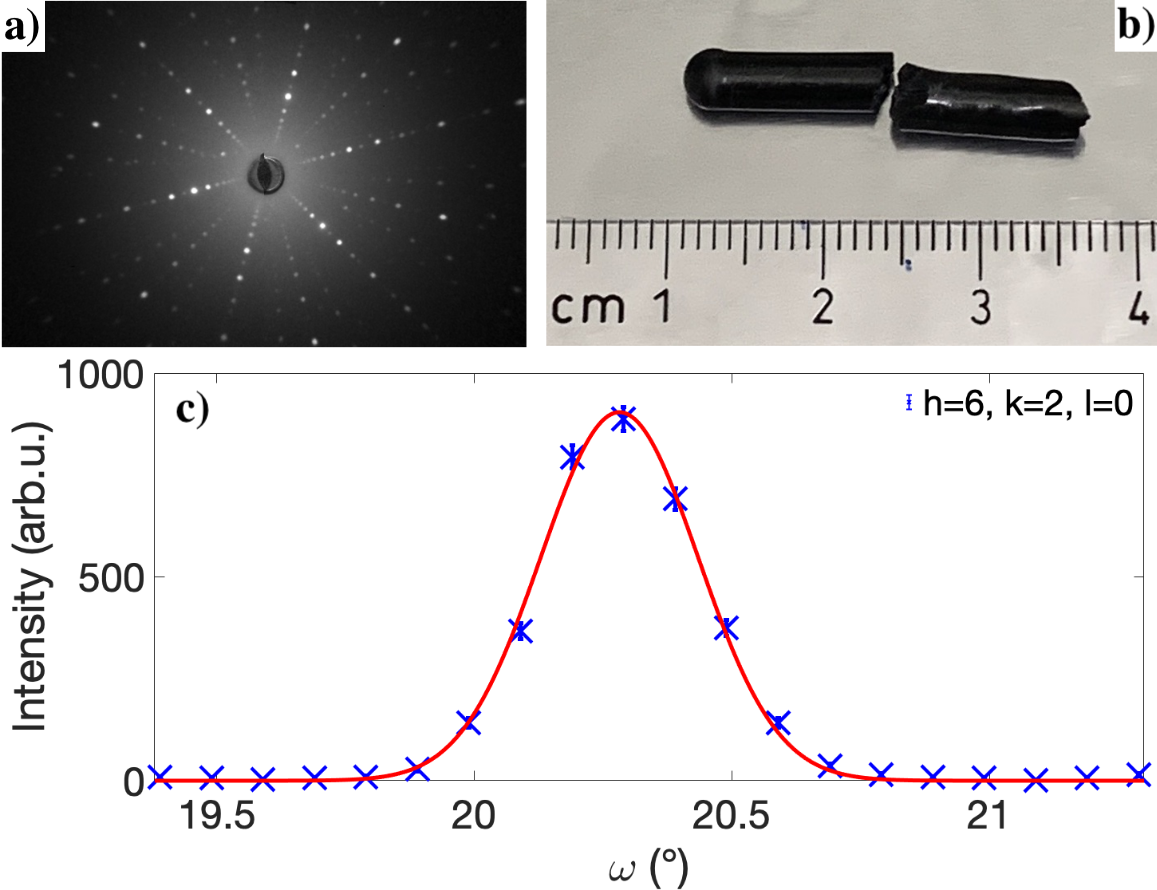}
\centering
\caption{(\textbf{a}) X-ray Laue pattern from our \cehf~single-crystal. (\textbf{b}) Single-crystal of \cehf~used for bulk and neutron diffraction experiments. (\textbf{c}) Rocking curve over the (620) reflection measured with Zebra on our single-crystal.}
\label{Picture.1}
\end{figure}

Polycrystalline samples of \cehf~were prepared via solid-state synthesis, using stoichiometric amounts of pre-annealed CeO$_2$ and HfO$_2$. The mixture was then thoroughly ground in a planetary ball mill for an hour. The recovered powder was then subject to several cycles of sintering and grinding until the observation of a single pyrochlore phase in X-ray diffraction patterns (Brucker D8 Advance Cu K$_{\alpha}$). Sintering was carried out at 1600~\degC~in a reductive atmosphere (He/H$_{2}$, 10\% H$_{2}$) for 150 hours in a High Temperature Horizontal Tube Furnace (HTRH, Carbolite Gero). A nonmagnetic, iso-structural, reference sample, \lahf, was prepared following a similar method although the sintering took place in air at 1300~\degC. Part of the \cehf~powder was subsequently used as starting material for single-crystal growth using the optical floating zone technique. The growth was performed at the Paul Scherrer Institut using the high pressure, high temperature optical-image travelling solvent floating zone furnace (HKZ by Scidre GmbH). The crystallinity and the phase purity of the as-grown single-crystal (Fig.~\ref{Picture.1}(\textbf{b})) were checked using both a X-ray Laue camera (Fig.~\ref{Picture.1}(\textbf{a})) and X-ray diffraction on a powdered fragment. Two slices with $[1\bar{1}0]$ normal, of 12.35~mg and 12.6~mg respectively, were prepared for subsequent macroscopic measurements. The rest of the crystal was used for neutron diffraction experiments.

\begin{figure*}[!ht]
\includegraphics[scale=0.35]{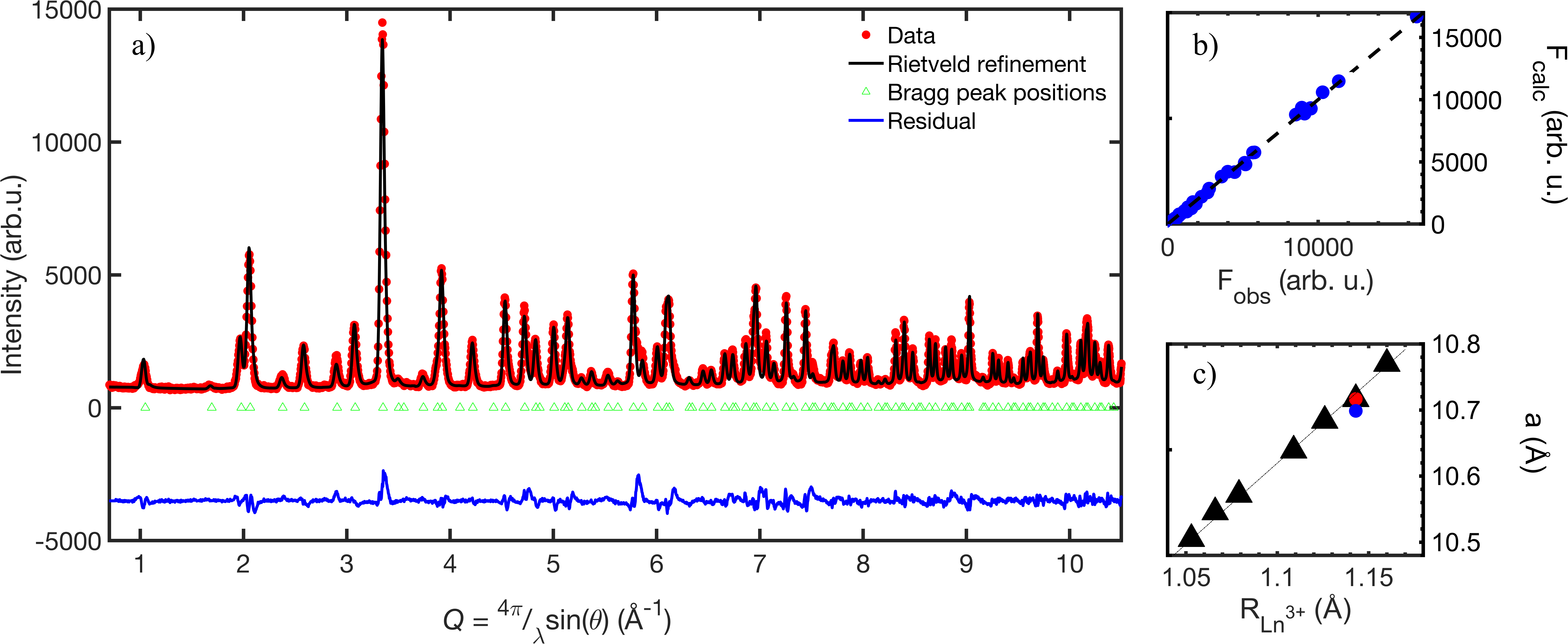}
\centering
\caption{(\textbf{a}) Rietveld refinement of high-resolution neutron diffraction pattern collected on the starting powder at 1.5~K. The model used corresponds to a perfect pyrochlore structure with additional oxygen ions located at the 8\textit{a} crystallographic site (Table~\ref{Table3}). (\textbf{b}) Nuclear refinement of our single-crystal neutron diffraction data using a similar model (Table~\ref{Table4}). (\textbf{c}) Relationship between the lattice parameter of rare-earth pyrohafnates and the ionic radii of trivalent rare-earth ions. Our values of the lattice parameter for the single-crystal and the starting powder are highlighted in red and blue, respectively. Only the compounds crystallizing in a pyrochlore structure were included.}
\label{Fig.1}
\end{figure*}

Thermogravimetric measurements were performed on a NETZSCH STA 449C thermogravimetric (TG) analyzer. The temperature was ramped up to 1000~\degC~at a rate of 3~\degC/min with an intermediate step of 3 hours at 400~\degC. The experimental chamber was subject to an oxygen flow during the whole process.
\newline
The magnetic response of the \cehf~polycrystalline sample was measured in a temperature window ranging from 1.8 to 375~K in an applied magnetic field of 0.1~T using a Quantum Design MPMS-XL superconducting quantum interference device (SQUID) magnetometer. In addition, low-temperature magnetization and $ac$-susceptibility measurements were made on polycrystalline samples as a function of temperature and field, from $T$~=~0.07 to 4~K and from $\mu_0H$~=~0 to 8~T, using SQUID magnetometers equipped with a miniature dilution refrigerator developed at the Institut N\'eel-CNRS Grenoble~\cite{Paulsen01}.

Heat capacity data, between 0.4 and 20~K, were obtained using a Quantum Design physical properties measurement system (PPMS) at the Institut N\'eel-CNRS Grenoble.

X-ray absorption spectroscopy (XAS) measurements of \cehf~(starting powder and powdered single crystal), \cesn$ $ and Ce$_{2}$Sn$_{2}$O$_{8}$~at the Ce L3-edge (5723~eV) were performed at the SuperXAS beamline~\cite{QuickEXAFS} of Swiss Light Source. The beam originating from a 2.9~T superbend magnet was collimated using a Si coated mirror at 2.9~mrad. The subsequent beam was monochromatised using a liquid nitrogen cooled Si 111 monochromator oscillating with a 1~Hz repetition rate. The beam was then focused onto the sample with a beam Rh-coated double focussing mirror with a spot size of 1~mm x 0.2~mm (HxV). Transmission geometry measurements using 1/2~bar He 1/2~bar N$_{2}$ filled ion chambers were performed for 5 minutes per sample. Energy calibration was performed using a simultaneously measured Cr foil. The raw data were processed using the {\it ProQEXAFS} software~\cite{proexafs} and the linear combination analysis of the absorption spectra was performed by means of the ATHENA software~\cite{Athena}.
\newline

The crystal-electric field states of the \ce~ions were probed using inelastic neutron scattering on the MAPS spectrometer~\cite{MAPS} at the ISIS Neutron and Muon Facility (Harwell, England). Large polycrystalline samples of \cehf~(18.3~g) and \lahf~(17~g) were inserted into cylindrical aluminium cans in an annular geometry, mounted on the cold head of a He closed-cycle refrigerator and measured using incident energies $E_i$ = 200~meV and $E_i$ = 500~meV at $T$ = 7~K. Data obtained with \lahf~were used to subtract lattice contributions to the \cehf~data.

Neutron powder diffraction was performed on HRPT~\cite{HRPT} (SINQ, Paul Scherrer Institut, Villigen, Switzerland) using a wavelength of 1.155~\AA. The 20~g \cehf~powder sample was placed in a vanadium can and cooled down using an Orange Cryostat for measurements at 1.5~K.
Single-crystal neutron diffraction experiments were carried out on Zebra (SINQ, Paul Scherrer Institut, Villigen, Switzerland) with an incident wavelength of 1.178~\AA ~(Fig.~\ref{Picture.1}(\textbf{c})). The nuclear structure was studied at room temperature, using a Eulerian cradle. All the neutron diffraction data were analyzed using the FullProf Suite software~\cite{FullProf_93} and the resulting crystallographic parameters are summarized in Appendix \ref{refinements}.
\section{Results and discussion}
\subsection{Synthesis and characterization}

As for oxides in general, stabilization of pyrochlores hosting \ce~ions is known to be rather difficult~\cite{CHO_CZO}. In the case of the hafnate variant, it requires very clean reducing conditions, which are particularly difficult to achieve because of the high annealing temperatures (1600~\degC) needed to form the compound. A large amount of polycrystalline Ce$_2$Hf$_2$O$_7$~was prepared via solid state synthesis, as detailed in the previous section. X-ray powder diffraction reflected a well ordered $Fd\bar{3}m$ pyrochlore phase with a lattice parameter of 10.7168(2)~\AA~at room temperature.

This value is slightly smaller than expected, based on the relationship between the lattice parameter of rare-earth pyrohafnates~\cite{KARTHIK2012168} and the ionic radii of trivalent rare-earth ions~\cite{ShannonRadii} (see Fig.~\ref{Fig.1}(\textbf{c})). A deviation can be explained by the presence of non-magnetic \cebis$ $ ions, which have a smaller ionic radius compared to \ce. Tetravalent cerium defects would also be accompanied by extra oxygen ions, so as to ensure electric neutrality. The precise oxygen stoichiometry of different pyrochlore materials was shown to be an important parameter impacting the magnetic properties of some compounds. For instance, it can select different ground-states as in Yb$_2$Ti$_2$O$_7$~\cite{Bowman} and Tb$_{2+x}$Ti$_{2-x}$O$_7$~\cite{TTO_Takatsu} or modify the excitations as seen in \dyti~\cite{Sala}.

A ground fragment of single-crystal was subject to the same analysis, yielding a further reduced lattice parameter of 10.6988(1)~\AA, suggesting a higher concentration of defects. In both cases, laboratory X-ray diffraction could not resolve the excess of oxygen, thus motivating the use of other techniques, as described hereafter.

\begin{figure}[!hb]
\includegraphics[width=\linewidth]{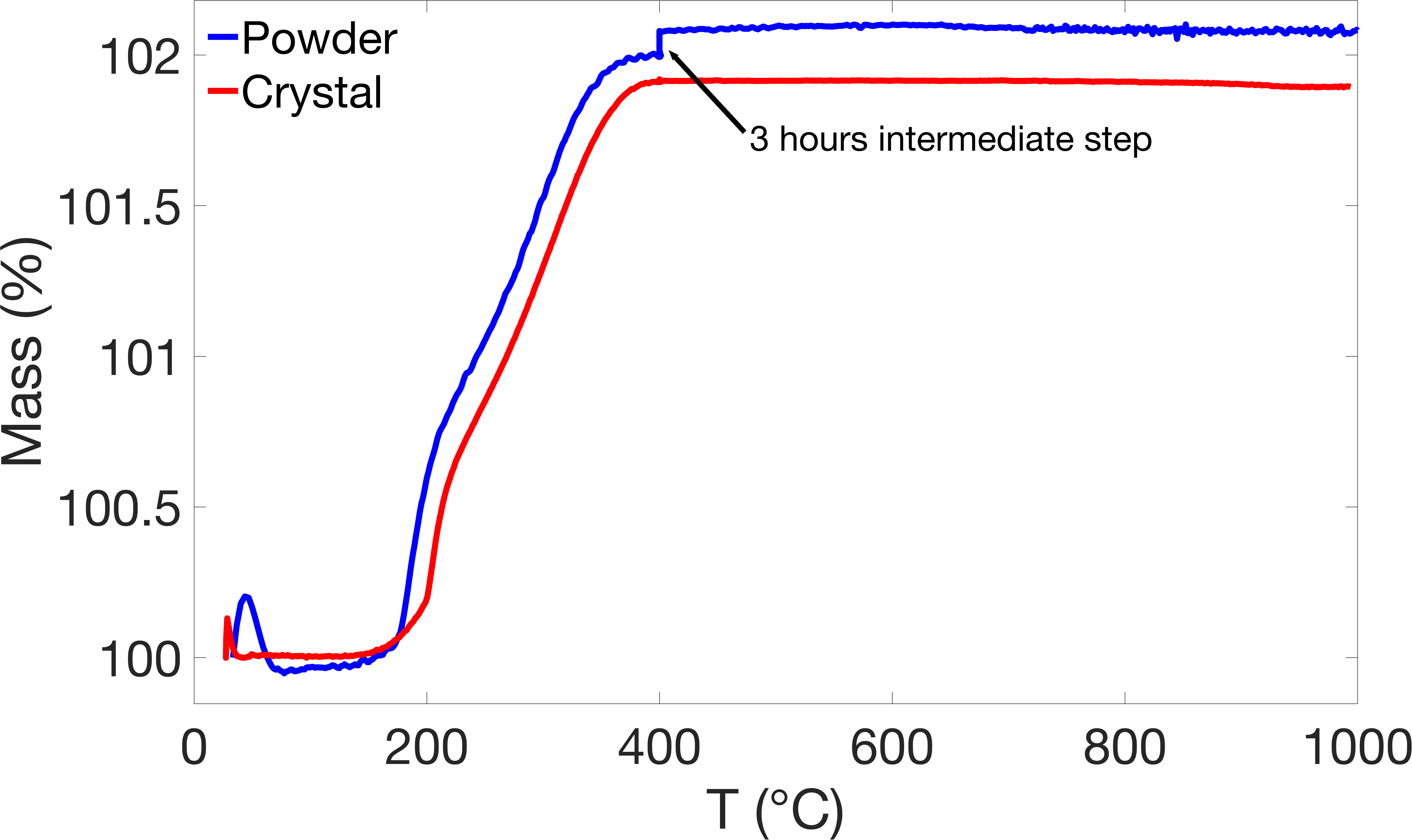}
\centering
\caption{Thermogravimetric measurement of our starting powder and powdered crystal of \cehf. The obtained mass gain is found to be $2.077\%\pm0.007$ and $1.190\%\pm0.001$, respectively. According to the reaction \cehf~+ (1 - $\delta$)/2 O$_{2}$ $\rightarrow$ 2 CeO$_{2}$ + 2 HfO$_{2}$, this mass gains correspond to $\delta_{powder} = 0.027\pm0.007$ and $\delta_{crystal} = 0.105\pm0.010$. X-ray diffraction of the products of the reactions corroborates the above chemical equation.}
\label{Fig.2}
\end{figure}

\begin{figure}[!hb]
\includegraphics[width=\linewidth]{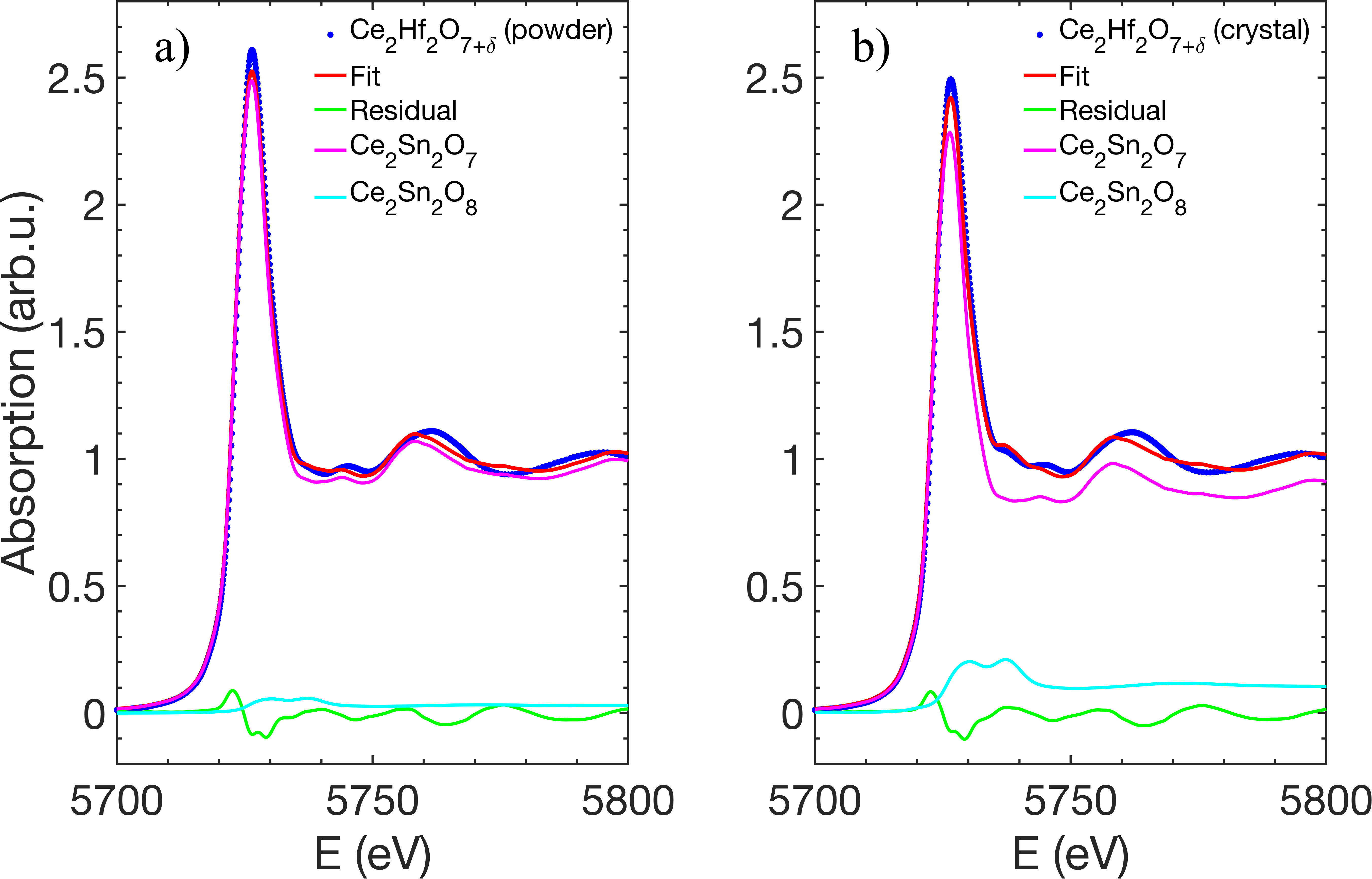}
\centering
\caption{X-ray absorption spectra of \cehf~at the Ce L3-edge measured on samples of starting powder (\textbf{a}) and powdered crystal (\textbf{b}). The signal was fitted using a linear combination of \cesn~(in pink) and Ce$_{2}$Sn$_{2}$O$_{8}$~(in cyan) spectra measured in identical conditions. The difference in height of the absorption edge is presumably coming from grain size effects.}
\label{Fig.3}
\end{figure}

Using Rietveld refinement, the neutron diffraction pattern of the \cehf~powder at 1.5~K (Fig.~\ref{Fig.1}(\textbf{a})) could be reproduced assuming a typical pyrochlore structure~\cite{PyroStr1,PyroStr2} together with the introduction of a small amount of oxygen atoms on the normally empty (8\textit{a}) crystallographic site (Appendix \ref{refinements}. Table~\ref{Table3}). The presence of such defects confirms the deviation from the nominal oxygen stoichiometry and implies that some cerium ions are indeed in a nonmagnetic tetravalent state. From Rietveld refinement, the amount of \cebis~defects is estimated to be around 6$\pm$1\%. However, the weak occupancy of the (8\textit{a}) site is rather unstable during the refinement. For this reason other techniques were employed to improve our estimate of the defect density. 

\begin{table}[t]
\begin{ruledtabular}
\begin{tabular}{ccc}
                  &   Starting powder   & Single-crystal     \\ \hline
Neutron diffraction   &      $6\pm 1 \%$     & $7 \pm 2 \%$             \\
Thermogravimetry     &    $2.7 \pm 0.7 \%$      &        $10.5 \pm 1 \%$      \\
XAS        &      $3 \pm 0.3 \%$     &    $11 \pm 0.3 \%$ 
\end{tabular}
\end{ruledtabular}
\caption{\label{Table1} Estimated amount of oxygen defects obtained from different methods on both powder and single-crystal samples.}
\end{table}

\begin{figure*}[]
	\begin{subfloat}
            \centering
            \includegraphics[scale=0.25]{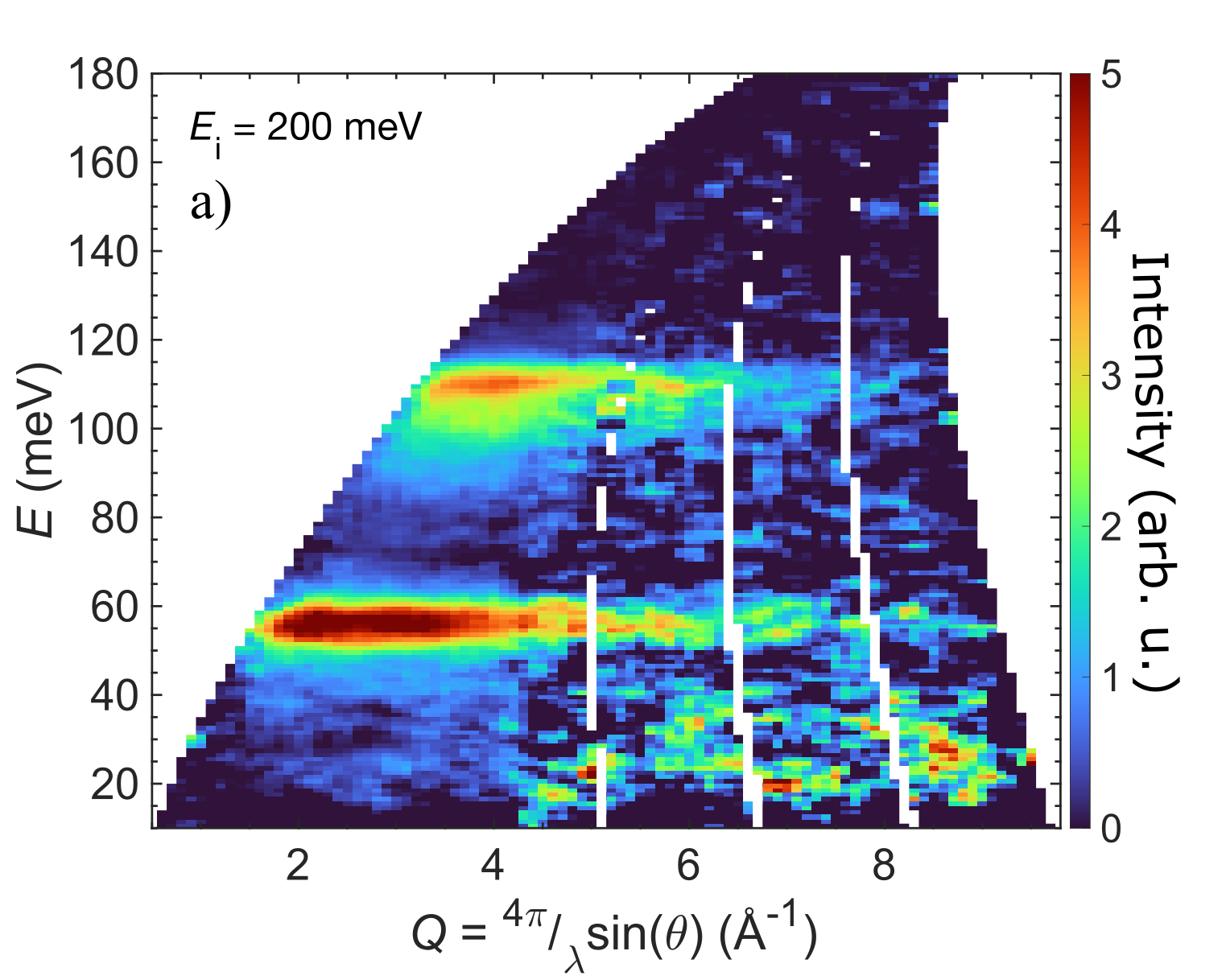}
        \end{subfloat}
        \begin{subfloat}
            \centering 
            \includegraphics[scale=0.25]{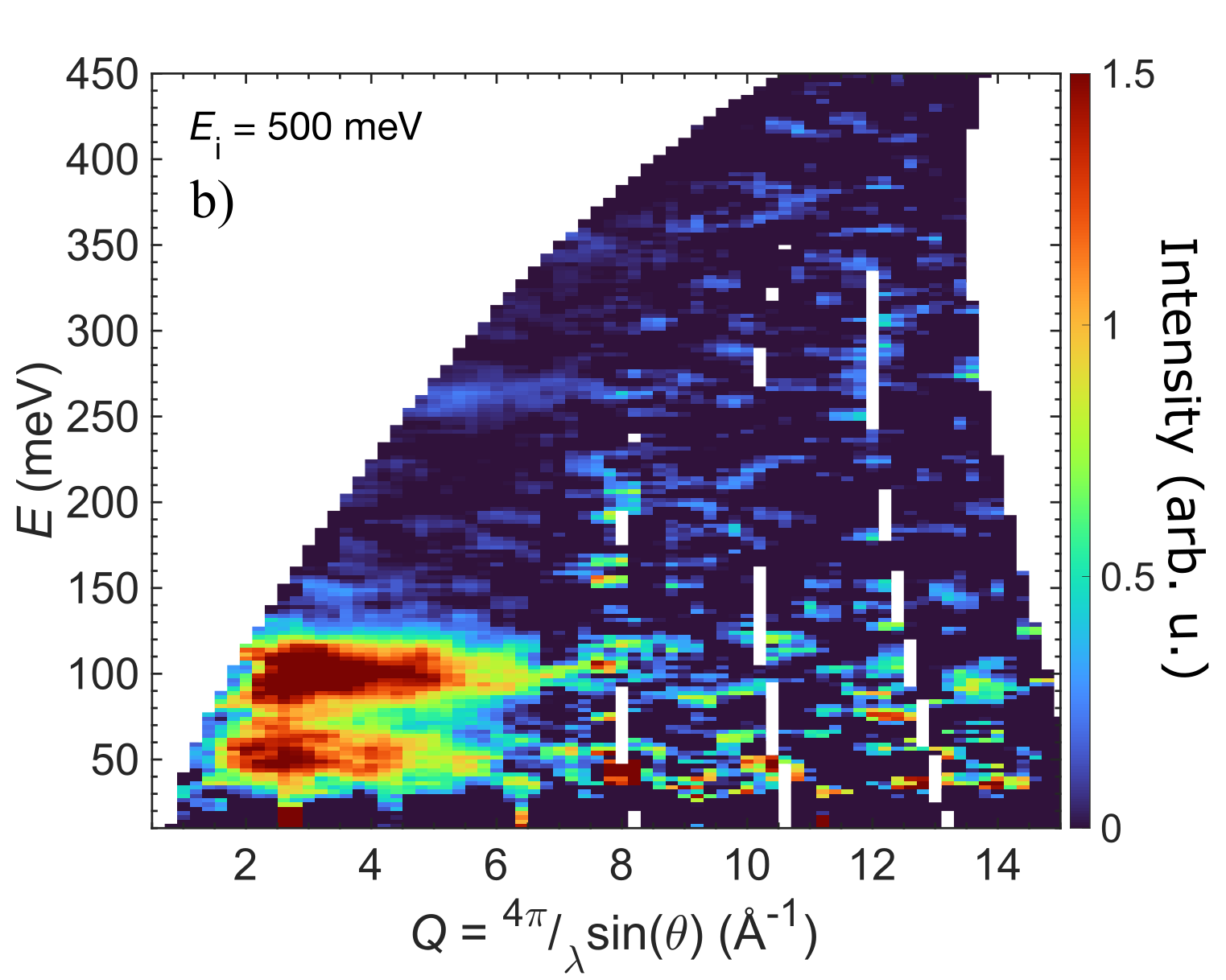}
        \end{subfloat}
        \vskip\baselineskip
        \begin{subfloat}
            \centering 
            \includegraphics[scale=0.175]{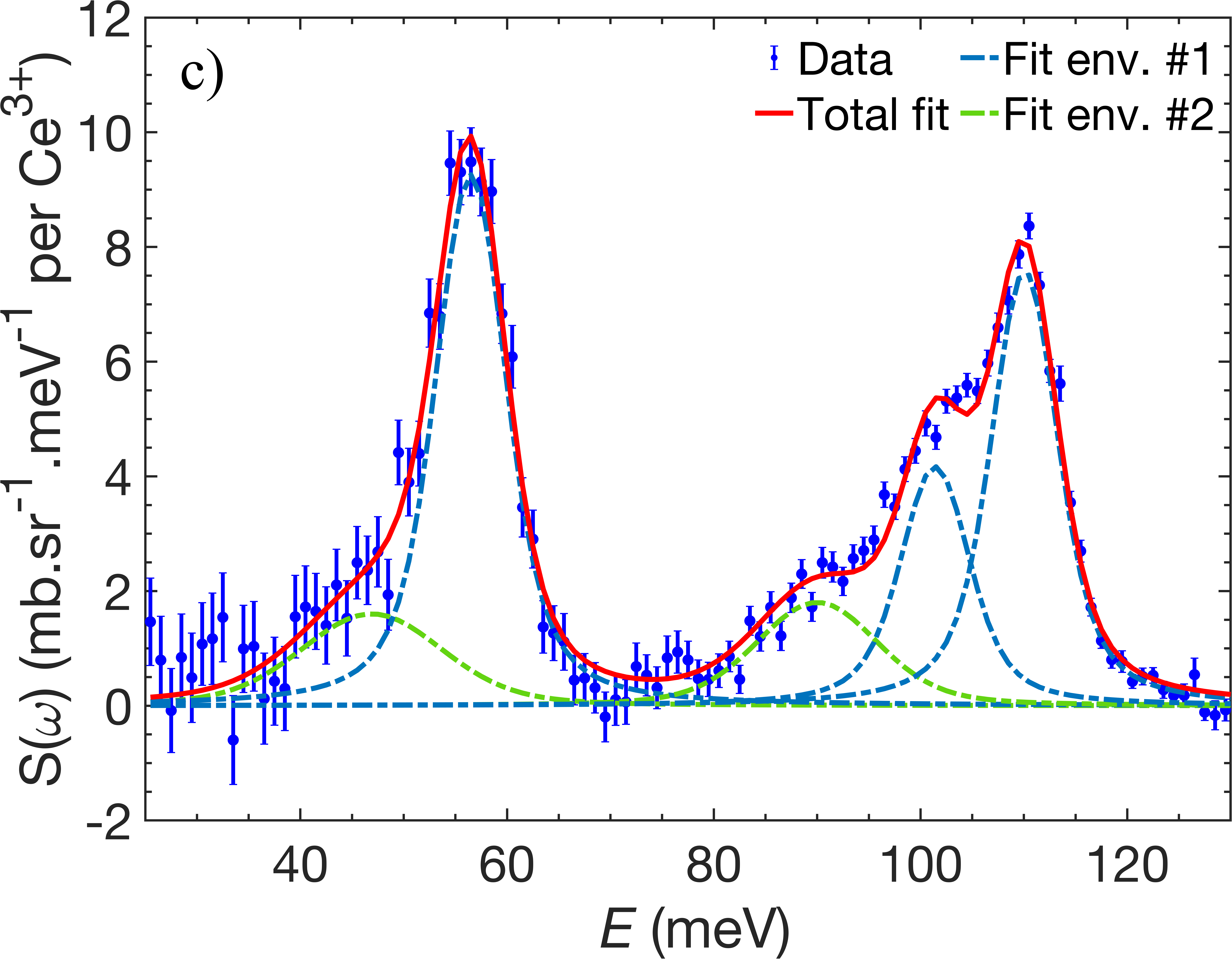}
        \end{subfloat}
        \begin{subfloat} 
            \centering 
            \includegraphics[scale=0.175]{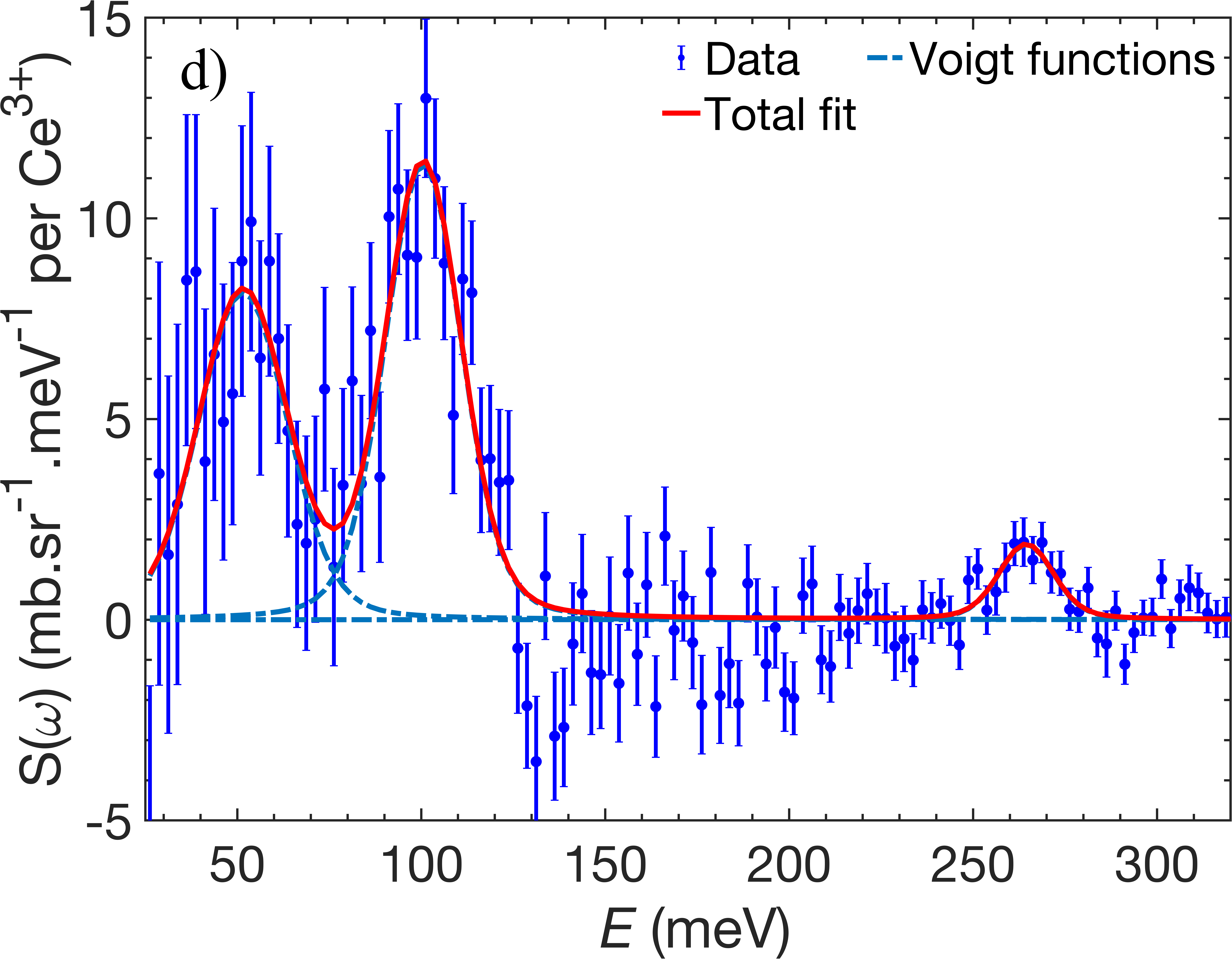}
        \end{subfloat}
\centering
\caption[]{(\textbf{a},\textbf{b}) Inelastic scattering neutron spectra measured using incident energies of 200~meV and 500~meV, respectively. Phonon contributions were accounted for by subtraction of the spectrum measured on the isostructural \lahf~compound. (\textbf{c},\textbf{d}) Constant-Q cut in the INS data above, corrected for magnetic form factor and integrated between 3.2~\AA$^{-1}$~and 4.5~\AA$^{-1}$ ($E_{i}$~=~200~meV) and 3.45~\AA$^{-1}$~and 7.4~\AA$^{-1}$ ($E_{i}$~=~500~meV). Voigt functions were used to fit the observed excitations, as described in the text. Only one set of excitations could be identified in the high energy data due to the coarser resolution.} 
\centering
\label{Fig.4}
\end{figure*}

The raw powder was subject to thermogravimetry analysis following a similar procedure as was used in [\onlinecite{TOLLA:1999tq}]. The variation of sample weight is directly related to changes in oxygen stoichiometry, thus providing a robust estimate of the tetravalent cerium concentration. In our powder samples, the oxygen stoichiometry was found to be 7.027$\pm$0.007 (see Fig.~\ref{Fig.2}), which translates into 2.7$\pm$0.7~\% of \cebis~defects. 

Finally, we have directly confirmed the change in cerium oxidation state using XAS. To provide a meaningful analysis using this technique, we first measured samples that can be used as references. The isostructural \cesn$ $ was shown to hold 100\% of \ce$ $ to a good approximation~\cite{CSO_NatPhy}. It is thus an appropriate choice to evaluate the amount of trivalent cerium ions. On the other hand, we use  Ce$_{2}$Sn$_{2}$O$_{8}$, where the cerium are all tetravalent and experience a cubic \textit{O$_{h}$} point group symmetry, as a reference to evaluate the amount of \cebis. The \cehf~x-ray absorption spectrum is well reproduced using a linear combination of the \cesn~and Ce$_{2}$Sn$_{2}$O$_{8}$~spectra (Fig.~\ref{Fig.3}). The best fit for our \cehf~powders is reached with a 3$\pm$0.3 \% contribution of Ce$_{2}$Sn$_{2}$O$_{8}$, which is in good agreement with our quantification of the defects obtained from thermogravimetry. We note that measurements performed on samples stored in helium and in air lead to very similar results, indicating that the compound is stable in air, at least on the scale of several months. The latter significantly differs from observations made on \cezr, by other groups~\cite{CZO_CA} as well as by us, that the lattice constant and colour of cerium zirconate show substantial variations after exposure to air of only a few days. This may indicate important differences in the reactivity of the \ce~zirconates and hafnates, which is not anticipated since tetravalent Hf and Zr are in principle very similar, both sterically and chemically.

We now turn to estimates of the stoichiometry in single-crystals.  Using single-crystal neutron diffraction data (Fig.~\ref{Fig.1}(\textbf{b})), refinements yield an anticipated higher amount of defects (7$\pm$2\%, Appendix \ref{refinements}. Table~\ref{Table4}), most probably stabilized during the high-temperature growth. The thermogravimetric analysis was also performed on a ground piece of the \cehf~crystal. In this case, 10.54$\pm$1\% of the cerium cations appear to be in a tetravalent oxidation state  (Fig.~\ref{Fig.2}). XAS measurements also corroborate this result with an estimated 11$\pm$0.3\% of \cebis~defects as can be seen on Fig.~\ref{Fig.3}(\textbf{b}). A summary of the results obtained using different techniques can be found in Table~\ref{Table1}.

\subsection{Single-ion properties}

We report inelastic neutron scattering (INS) data probing the crystal-electric field splitting of \ce~in powders of \cehf$ $ at $T$~=~7~K (Fig.~\ref{Fig.4}(\textbf{a},\textbf{b})). Phonon contributions were removed by subtraction of the \lahf~data collected in an identical configuration. The resulting data were corrected considering the \ce~magnetic form factor. Using an incident energy of 200~meV, five CEF excited states could be observed, as can be seen on Fig.~\ref{Fig.4}(\textbf{c}). The two main peaks, located at  56~meV and 110~meV, are representative of the transitions from the ground-state doublet to the two excited Kramers doublets of the $J=5/2$ multiplet.

\begin{figure}[]
\includegraphics[scale=0.21]{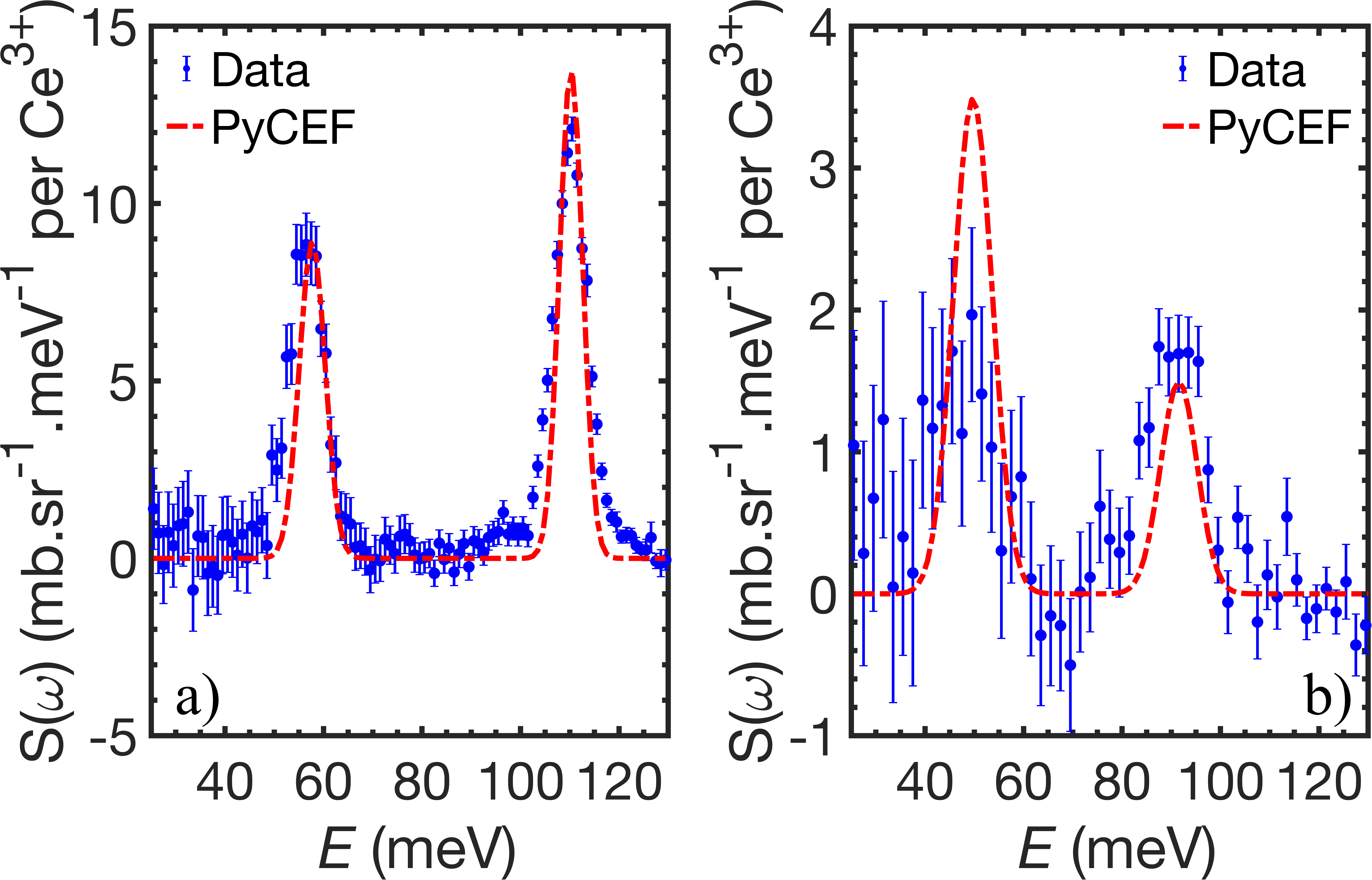}
\centering
\caption{Decomposed spectra with the CEF excitations coming from \ce$ $ experiencing a perfect D$_{3d}$ (\textbf{a}) and defective (\textbf{b}) environment, using an incident energy of 200~meV. The red lines represent the fits obtained with PyCrystalField using two different CEF Hamiltonians.}
\label{Fig.4e}
\end{figure}

\begin{figure*}[]
\includegraphics[scale=0.325]{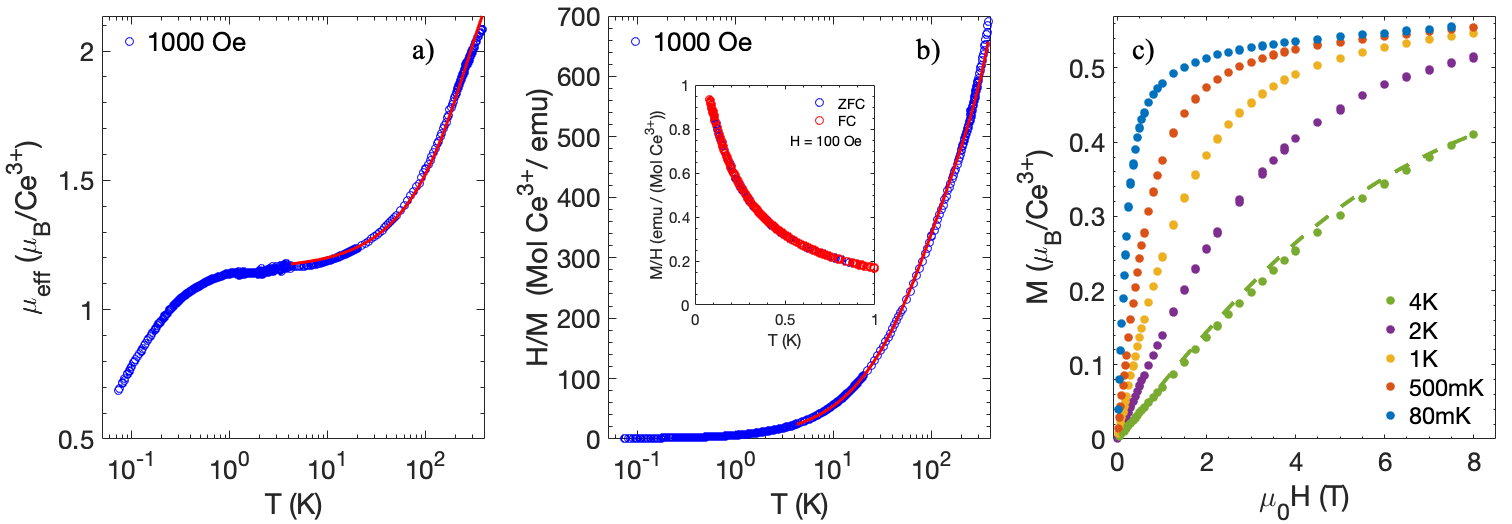}
\centering
\caption{(\textbf{a}) Effective magnetic moment obtained from the magnetization measured at 1000~Oe and also shown as the inverse magnetic susceptibility (\textbf{b}). In both (\textbf{a}) and (\textbf{b}), the blue circles are experimental data and the red line corresponds to our CEF model based on susceptibility and INS data analysis. The inset shows the magnetic susceptibility obtained under field-cooled (FC) and zero-field cooled (ZFC) conditions applying a field of 100~Oe. (\textbf{c}) $M$ vs $H$ measured at different temperatures on a powder sample. The dashed line is the powder-averaged magnetization curve at 4~K calculated from our PyCrystalField model.}
\label{Fig.5}
\end{figure*}

A third excitation visible around 100~meV is reminiscent of the additional CEF excitation observed in \cezr~\cite{CZO_CA} and is likely to stem from hybridization of the CEF with a phonon. Hybridized excitations of mixed magnetic and phononic character were observed in other pyrochlore magnets, such as in \przr~\cite{PZO_Raman}, \tbti~\cite{Fennell2014} as well as other \ce-based compounds~\cite{CeCuAl3, CeAuAl3, CeCuGa3}. From their weaker spectral weight, the two remaining excitations (around 47~meV and 90~meV) are most likely due to a second CEF environment related to interstitial oxygen atoms (O"(8\textit{a})). The estimated 3~\% of interstitial oxygens translates into 18~\% of the cerium sites experiencing a  disturbed crystallographic environment. This effectively lowers the symmetry with respect to the usual $D_{3d}$ point group symmetry at the pyrochlore's A site, down to the $C_1$ point group symmetry where all CEF parameters are allowed. However, a symmetry lowering does not lift any degeneracy as the three doublets of the $J=5/2$ multiplet of this Kramers ion are already maximally separated. Another measurement with incident energy of 500~meV allowed the observation of an additional excitation located at 264~meV (Fig.~\ref{Fig.4}(\textbf{d})), explained by a transition from the ground state to the first level of the $J=7/2$ multiplet.

The three major CEF excitations were fitted using Voigt functions, restricting their gaussian widths to the instrumental resolution (Fig.~\ref{Fig.4}(\textbf{c})). The two weaker excitations, stemming from the presence of impurities, could be fitted using the same function but with larger widths (Fig.~\ref{Fig.4}(\textbf{d})). This can be justified by a probable distribution of defective environments. The measured spectrum was then deconvoluted into two artificial spectra (Fig.~\ref{Fig.4e}(\textbf{a-b})), respectively corresponding to the perfect and defective CEF environments. The first one contains the two main peaks at 56 and 110~meV. The fitted Voigt of the extra peak at 100~meV was also added on top of the excitation around 110~meV to account for the observed splitting. However, its width was changed to match the experimental resolution at the corresponding energy. The second spectrum solely contains the weaker excitations attributed to cerium ions experiencing a defective environment. The two decomposed spectra were subsequently used to fit the crystal field parameters using the PyCrystalField~\cite{PyCEF} package. To do so, we defined a model with two crystal field Hamiltonians, both using the whole set of 14 intermediate-coupling basis states of the $J=5/2$ and 7/2 multiplets~\cite{Boothroyd1992}. This is particularly important for light rare-earth elements where the strength of the spin-orbit coupling competes with the crystal field. The first Hamiltonian is representative of \ce~in a perfect $D_{3d}$ symmetry, thus incorporating only six non-zero CEF parameters ($B_{2}^{0}$, $B_{4}^{0}$, $B_{4}^{3}$, $B_{6}^{0}$, $B_{6}^{3}$ and $B_{6}^{6}$). The second one includes \cebis~defects in the second coordination sphere of the \ce~ions as well as interstitial oxygens (8\textit{a}), allowing all CEF parameters to be non-zero. We began by fitting the CEF parameters to the observed transition energies in order to get reasonable starting values for the subsequent steps. We then included the intensities of these excitations observed in Fig.~\ref{Fig.4e}(\textbf{a}). The intensity of the excitation around 260~meV was not considered, because the present resolution does not allow to identify contributions from the two CEF environments. Nonetheless, both the positions and the intensities of the observed transitions are well accounted for by our model (Table~\ref{Table2} and Fig.~\ref{Fig.4e}(\textbf{a},\textbf{b})). We note that the fit results in a fraction of defective environment of 24~\%. This corresponds to 3.8~\% of \cebis, which is in the same range as the estimate obtained with both thermogravimetry and XAS. 

Taking our analysis a step further, we integrated the magnetic susceptibility measured on the powder sample to the fitting routine. The contributions from each environment were weighted based on the estimated amount of defects (3~\% of \cebis). The result of the fit is shown on Fig.~\ref{Fig.5}(\textbf{a-b}). From the entire CEF analysis, it follows that the wavefunction of the ground state doublet is a linear combination of the $|m_{J_{z}} = \pm 3/2 \rangle$ states, as was found in other cerium-based pyrochlores~\cite{CSO_PRL,CSO_NatPhy,CZO_CA,CZO_US}. This is consonant with a pseudo-spin carrying both dipolar and octupolar components~\cite{DO-PRB} as well as a strong Ising anisotropy. For the perfect CEF environment, we compared the results from PyCrystalField and SPECTRE~\cite{SPECTRE} using the same set of CEF parameters. Both programs returned quite similar expected neutron cross-sections and ground state wavefunctions. Results obtained with SPECTRE can be seen in Appendix~\ref{SPECTRE_benchmarking}.

We note that the number of allowed parameters of the second CEF environment compared to the number of observables does not enable to provide a reliable analysis. This limitation explains the rather poor fit of the defective CEF presented in Fig.~\ref{Fig.4e}(\textbf{b}), and is expected to contribute to the small deviation between the measured magnetic susceptibility and the calculation (see Fig.~\ref{Fig.5}(\textbf{a}-\textbf{b})). The values of CEF coefficients for the defective environment can be found in Appendix~\ref{CEF_2}. Although we cannot conclude on a precise composition of the wavefunction of the ground state doublet for the defective environment, the dipole-octupole nature of the latter appears to be robustly preserved as the $|m_{J_{z}} = \pm 3/2 \rangle$ states remain largely predominant.

\begin{table}[h]
\begin{ruledtabular}
\begin{tabular}{cccccc}
Degeneracy & $E_{obs}$    & $E_{calc}$  & $I_{obs}$ & $I_{calc}$ & CEF env.\# \\
\hline
2    & 0       & 0   & 	&    & 1 - 2       \\
2    & 46.9(1.2)  & 49.8  &	0.30(0.18) & 0.47 & 2           \\
2    & 56.5(0.3) & 57.5  &	1.00 & 1.00 & 1           \\
2    & 90.0(1.1)  & 91.7  & 0.29(0.10) & 0.19 & 2           \\
2    & 110.1(0.2) & 110.3 & 1.26(0.08) & 1.37 & 1   \\
2    & 264.4(0.1) &  280.6 &   0.14(0.03) & 0.48 & 1    
\end{tabular}
\end{ruledtabular}
\caption{\label{Table2} Observed and calculated crystal-field transition energies of \cehf~in the intermediate coupling scheme. The first column gives the level degeneracy which is always two here due to the Kramers nature of the \ce~ion. \textit{E$_{obs}$}~and \textit{E$_{calc}$}~are the observed and calculated crystal-field transition energies, with \textit{I$_{obs}$}~and \textit{I$_{calc}$}~the associated intensities (normalized to the intensity of the peak at 56.5~meV). The last column indicates to which model the transition belongs: 1 for the perfect D$_{3d}$ environment and 2 for the defective one. The intensity of the transition located at 110~meV is actually the sum of the intensities located at 101 and 110 meV, taking into account the observed splitting. The values of Stevens operators obtained from the fit were $B^0_2 = -0.820(188)$, $B^0_4 = 0.223(5)$, $B^3_4 = 1.773(53)$, $B^0_6 = -0.008(1)$, $B^3_6 = 0.074(2)$ and $B^6_6 = -0.075(2)$~meV for the first defined environment. The obtained ground-state wavefunction is $|\pm\rangle$ = 0.893$|^{2}F_{5/2}$,$\pm3/2\rangle$ $\mp$ 0.427$|^{2}F_{7/2}$,$\pm3/2\rangle$ + 0.134$|^{2}F_{7/2}$,$\mp3/2\rangle$ - 0.005$|^{2}F_{5/2}$,$\mp3/2\rangle$.}
\end{table}

The magnetization curves measured on the starting powder sample up to 8~T (Fig.~\ref{Fig.5}(\textbf{c})) saturate at roughly half the effective magnetic moment, which is expected for Ising moments on a pyrochlore lattice due to the important non-collinear local anisotropy~\cite{Bramwell:2000tc}. This strong local anisotropy is also consistent with the large energy gap to the first excited state observed in the INS data.

\begin{figure}[!h]
\includegraphics[scale=0.42]{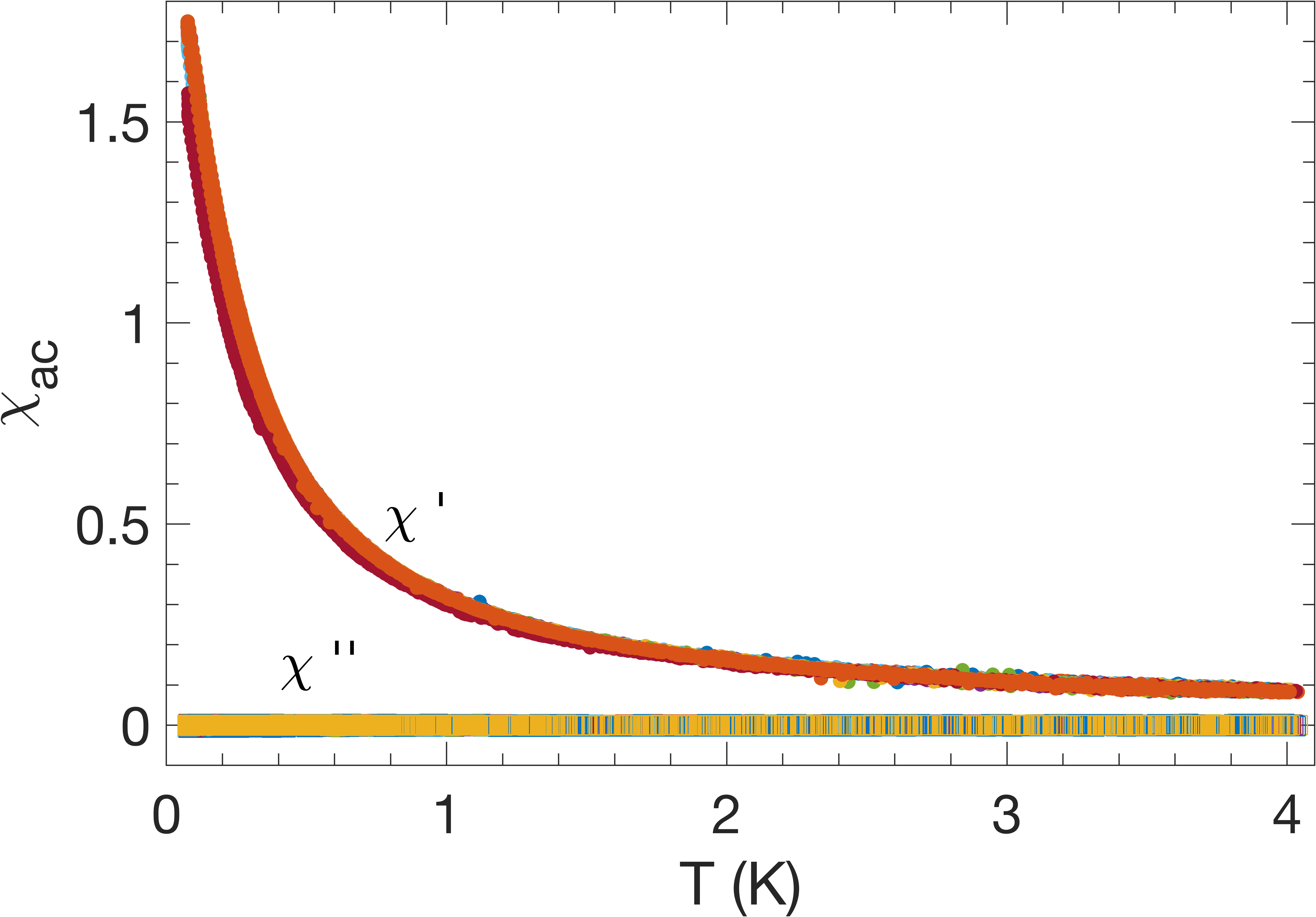}
\centering
\caption{(\textbf{a-b}) \textit{ac}-susceptibility measured on \cehf~powder down to 0.08~K with frequencies ranging from 0.11~Hz to 211~Hz.}
\label{Fig.6}
\end{figure}

\begin{figure}[!h]
\includegraphics[scale=0.28]{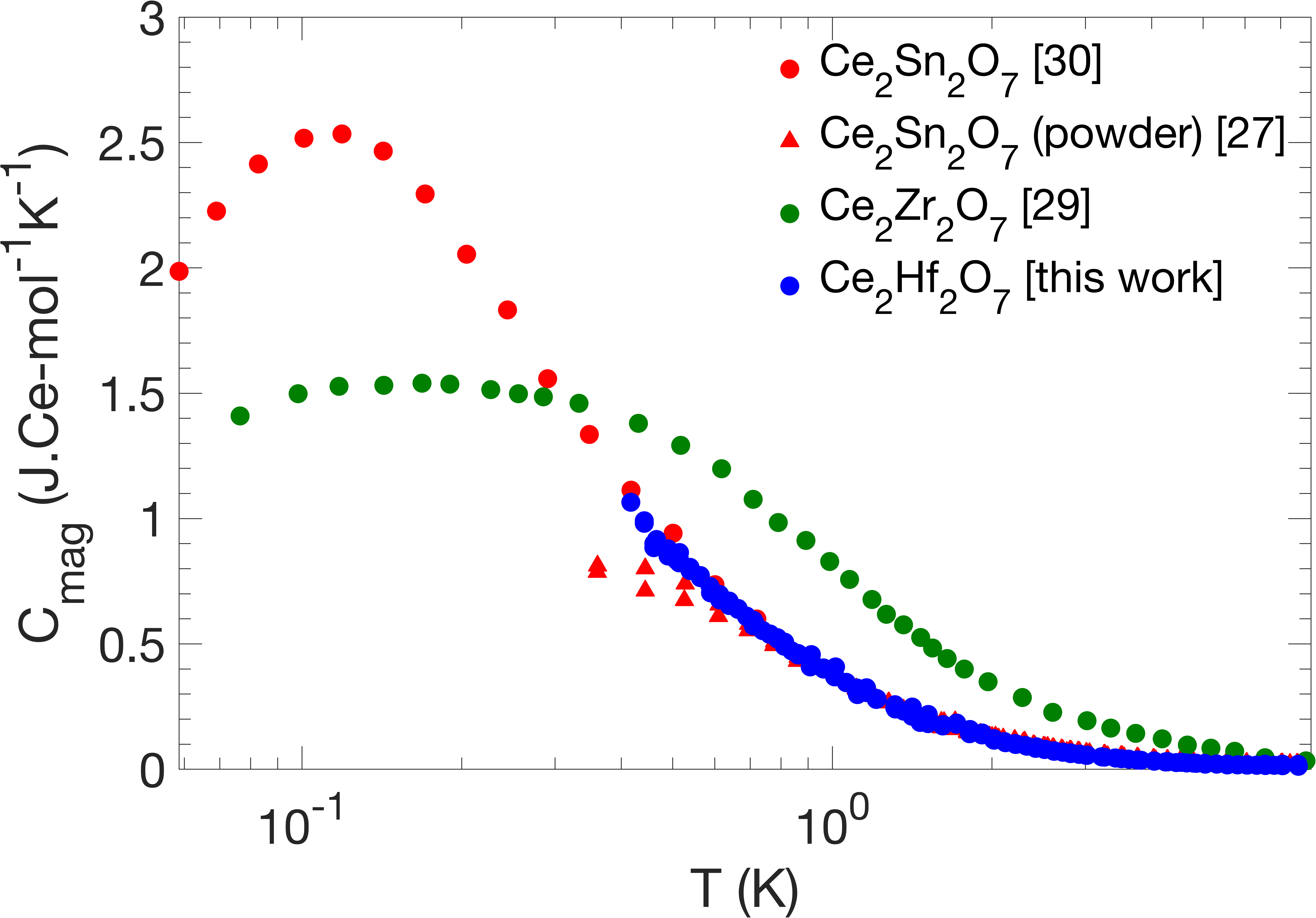}
\centering
\caption{Magnetic contribution to the heat capacity measured on a piece of \cehf~single-crystal and compared to similar data on other cerium pyrochlores. The magnetic contribution to the heat capacity of \cehf~was obtained by subtracting lattice contributions using data for \lahf~that were obtained in the same way. The experimental data for \cesn~and \cezr~are reproduced from references~\cite{CSO_crystal,CSO_NatPhy}~and~\cite{CZO_US}, respectively. We note that the discrepancy observed between the magnetic contributions to the heat capacity of \cesn~from references~\cite{CSO_NatPhy}~and~\cite{CSO_crystal}~only concerns few data points from~\cite{CSO_NatPhy}~that were measured between 0.5~K and the lowest temperature achieved (0.36~K). This is unlikely to reflect different correlated ground states between \cesn~samples used in~\cite{CSO_NatPhy}~and~\cite{CSO_crystal}~but rather results from difficulties in measuring heat capacity on powder samples below 0.5~K.}
\label{Fig.Cp}
\end{figure}

\subsection{Low-temperature correlated state}

The effective magnetic moment of the ground state doublet calculated from the crystal field ($\mu^{CEF}_{eff}$ = 1.18 $\mu_{B}$) compares very well with the value measured at temperatures between 1 K and 10 K ($\mu^{exp}_{eff}$ = 1.18(1) $\mu_{B}$, Fig.~\ref{Fig.5}(\textbf{a})), and is also close to the value observed in \cesn~\cite{CSO_PRL,CSO_NatPhy}. Upon cooling below about 1~K, a drop of the effective magnetic moment is observed (see Fig.~\ref{Fig.5}(\textbf{a})). The same behavior was found in \cesn~\cite{CSO_PRL,CSO_NatPhy}, where it was interpreted as a consequence of the evolution of the mixing of the wavefunctions of both elements of the doublet under the influence of dominant octupole-octupole interactions. 
Magnetic susceptibility measured on the polycrystalline sample did not display any sign of long-range order down to 0.08~K, nor any zero-field-cooled/field-cooled effects (see inset of Fig.~\ref{Fig.5}(\textbf{b})). Furthermore, \textit{ac}-susceptibility data do not exhibit any sizeable $\chi''$ response at frequencies ranging from 0.11~Hz up to 211~Hz (as can be seen with the flat signal at zero in Fig.~\ref{Fig.6}). This indicates that spins evade freezing down to the lowest temperatures achieved here.

Heat capacity was measured on a piece of \cehf~single-crystal down to 0.4~K in zero field. Lattice contributions were subtracted using identical measurements carried out on the isostructural compound \lahf~and the resulting magnetic contribution to the heat capacity can be seen on Fig.~\ref{Fig.Cp}. The onset of correlations appears as a strong increase of $C_{mag}$ when the temperature decreases below 2~K. We anticipate that it would result in a maximum at lower temperature, outside of the temperature range covered in this measurement. 
In order to compare the magnetic contributions to the heat capacity with other \ce~pyrochlores, we also reproduce in Fig.~\ref{Fig.Cp} literature data for \cesn~\cite{CSO_crystal,CSO_NatPhy}~and \cezr~\cite{CSO_NatPhy}. Remarkably, the results obtained for \cehf~is very similar to data reported for both powders~\cite{CSO_NatPhy}~and single-crystals~\cite{CSO_crystal}~of \cesn, while \cezr~\cite{CZO_US} shows a significantly different behavior, with a rise of magnetic contribution starting up at higher temperatures and being spread on a wider temperature range. Overall this suggests that correlations in various \ce-based pyrochlores, although similar, may have slightly different energy scales.

\section{Conclusions}
To summarize, we present a study of Ce$_2$Hf$_2$O$_{7}$, an alternative \ce-based pyrochlore having magnetic properties similar to that of its stannate and zirconate counterparts. We start by establishing the concentration of defects in the materials under investigation using several techniques, yielding around 3\% of non-magnetic tetravalent cerium in the polycrystalline sample and 11\% in the single-crystal. The understanding of these defects plays an important role in the detailed crystal-electric field analysis. The calculated \ce~ground-state wavefunction supports a dipole-octupole ground state doublet that is thermally isolated at the temperatures relevant for correlated physics in such materials. 
The magnetic susceptibility extracted from the CEF analysis is in good agreement with experimental data measured at temperatures between 375~K and 1~K. Below this temperature, the evolution of the magnetic susceptibility is reminiscent of the progressive reduction of the dipole component of the magnetic moment without any sign of long-range magnetic order or spin freezing that was also observed in \cesn. Together with the similar single-ion ground state wavefunction, this suggests the presence of the same type of correlated phase in \cehf.
Therefore, Ce$_2$Hf$_2$O$_{7}$ is a promising system to further explore the exotic properties of an octupolar quantum spin ice. Ultimately, large single-crystals of cerium pyrochlores should serve further experiments such as momentum-resolved inelastic neutron scattering, as well as thermodynamic measurements and neutron diffraction under specific directions of applied magnetic field in order to test recent theoretical predictions for the octupolar quantum spin ice~\cite{Placke}.
\newline
\newline
\newline
\section{Acknowledgements}
We acknowledge funding from the Swiss National Science Foundation (projects No. 200021\_179150). Low temperature magnetization experiments were funded by the European Commission under grant agreement no. 824109 European 'Microkelvin Platform'. Experiments at the ISIS Neutron and Muon Source were supported by a beamtime allocation from the Science and Technology Facilities Council (10.5286/ISIS.E.RB2010672). This work is based also on experiments performed at the Swiss spallation neutron source SINQ, Paul Scherrer Institute, Switzerland. We thank C. Paulsen for the use of his magnetometers and A. Scheie for his help with the PyCrystalField package. We thank A. Bhardwaj, H. Changlani, H. Yan, A. Nevidomskyy and M. Kenzelmann for fruitful discussions.
\newline
\newline
\appendix
\section{Results of structural refinements} \label{refinements}

The crystallographic parameters resulting from the Rietveld refinement of the neutron powder diffraction data and least-square refinement of the single-crystal neutron diffraction data are given in Table~\ref{Table3} and Table~\ref{Table4}, respectively.

\begin{table*}[!h]
\begin{ruledtabular}
\begin{tabular}{cccccccc}
a = 10.6938(6)\AA                     &   \textit{T} = 1.5 K   & x         &           & y         &           & z         & occ     \\ \hline
Ce (16\textit{d})                                          &           & 0.5       &           & 0.5       &           & 0.5       & 1       \\
Hf (16\textit{c})                                          &           & 0         &           & 0         &           & 0         & 1       \\
O (48\textit{f})                                           &           & 0.375     &           & 0.375     &           & 0.375     & 1       \\
O' (8\textit{b})                                           &           & 0.333(2)  &           & 0.125     &           & 0.125     & 1       \\
O'' (8\textit{a})                                          &           & 0.125     &           & 0.125     &           & 0.125     & 0.06(1) \\ \hline
ADPs in \AA$^2$ & $U_{11}$ & $U_{22}$ & $U_{33}$ & $U_{12}$ & $U_{13}$ & $U_{23}$ &         \\
Ce (16\textit{d})                                          & 0.00116   & 0.00116   & 0.00116   & 0.00047   & 0.00047   & 0.00047   &         \\
Hf (16\textit{c})                                          & 0.00075   & 0.00075   & 0.00075   & -0.00009  & -0.00009  & -0.00009  &         \\
O (48\textit{f})                                           & 0.00167   & 0.00125   & 0.00125   & 0         & 0         & 0.00000   &         \\
O' (8\textit{b})                                           & 0.00166   & 0.00126   & 0.00126   & 0         & 0         & 0.00040         &         \\
O'' (8\textit{a})                                          & 0.00465   & 0.00465   & 0.00465   & 0         & 0         & 0         &         
\end{tabular}
\end{ruledtabular}
\caption{\label{Table3} Structural parameters obtained from Rietveld refinement of our starting polycrystalline sample of \cehf~measured at 1.5 K (space group \textit{Fd$\bar{3}$m}, origin choice 2): R$_{Bragg}$ = 6.68; R$_{F}$ = 4.49. The lattice parameters, the \textit{x} coordinate of the oxygen (48\textit{f}), the occupation of the oxygen (8\textit{a}) as well as all the anisotropic temperature factors were refined.}
\end{table*}

\begin{table*}[!h]
\begin{ruledtabular}
\begin{tabular}{cccccccc}
a = 10.6988(1)\AA                     &   T = 297 \textit{K}   & x                    & y                    & z   &U$_{iso}$      & occ     \\ \hline
Ce (16\textit{d})                                          &           & 0.5                  & 0.5                  & 0.5   & 0.51287   & 1       \\
Hf (16\textit{c})                                          &           & 0                    & 0                    & 0    & 0.20366     & 1       \\
O (48\textit{f})                                           &           & 0.375                & 0.375                & 0.375  & 0.69443   & 1       \\
O' (8\textit{b})                                           &           & 0.331(3)             & 0.125                & 0.125  & 0.61752   & 1       \\
O'' (8\textit{a})                                          &           & 0.125                & 0.125                & 0.125  & 0.83292   & 0.07(2) \\ 
\end{tabular}
\end{ruledtabular}
\caption{\label{Table4}Structural parameters obtained from least-square refinement of neutron diffraction on our crystal of \cehf~measured at room temperature using a Eulerian cradle (space group \textit{Fd$\bar{3}$m}, origin choice 2): R$_{F}$ = 4.04. The \textit{x} coordinate of the oxygen (48\textit{f}), the occupation of the oxygen (8\textit{a}) as well as all the anisotropic temperature factors were refined. The value of the lattice parameter was obtained via X-ray diffraction, at room temperature, on a ground fragment of crystal.}
\end{table*}

\section{Defective crystal electric field environment} \label{CEF_2}
As mentioned in the main text, the number of allowed parameters of the second CEF environment, the defective one, is too large compared to the number of observable CEF transitions. This part of the fit is thus underconstrained and the obtained fitting parameters are not very reliable. 
In the Stevens formalism, these CEF parameters are
$B_2^0=2.638$, $B_2^1=-0.152$, $B_2^2=-0.029$, $B_4^0=0.160$, $B_4^1=0.001$, $B_4^2=0.001$, $B_4^3=0.182$, $B_4^4=8.749e-05$, $B_6^0=-0.007$, $B_6^1=3.02e-06$, $B_6^2=-1.470e-05$, $B_6^3=0.038$, $B_6^4=-1.016e-05$, $B_6^5=-1.663e-05$ and $B_6^6=-0.025$~meV. The corresponding ground-state wavefunction can be found in table \ref{TableCEF2}. 

\section{Crystal-electric field calculation using SPECTRE} \label{SPECTRE_benchmarking}
\begin{figure}[!h]
\includegraphics[scale=0.21]{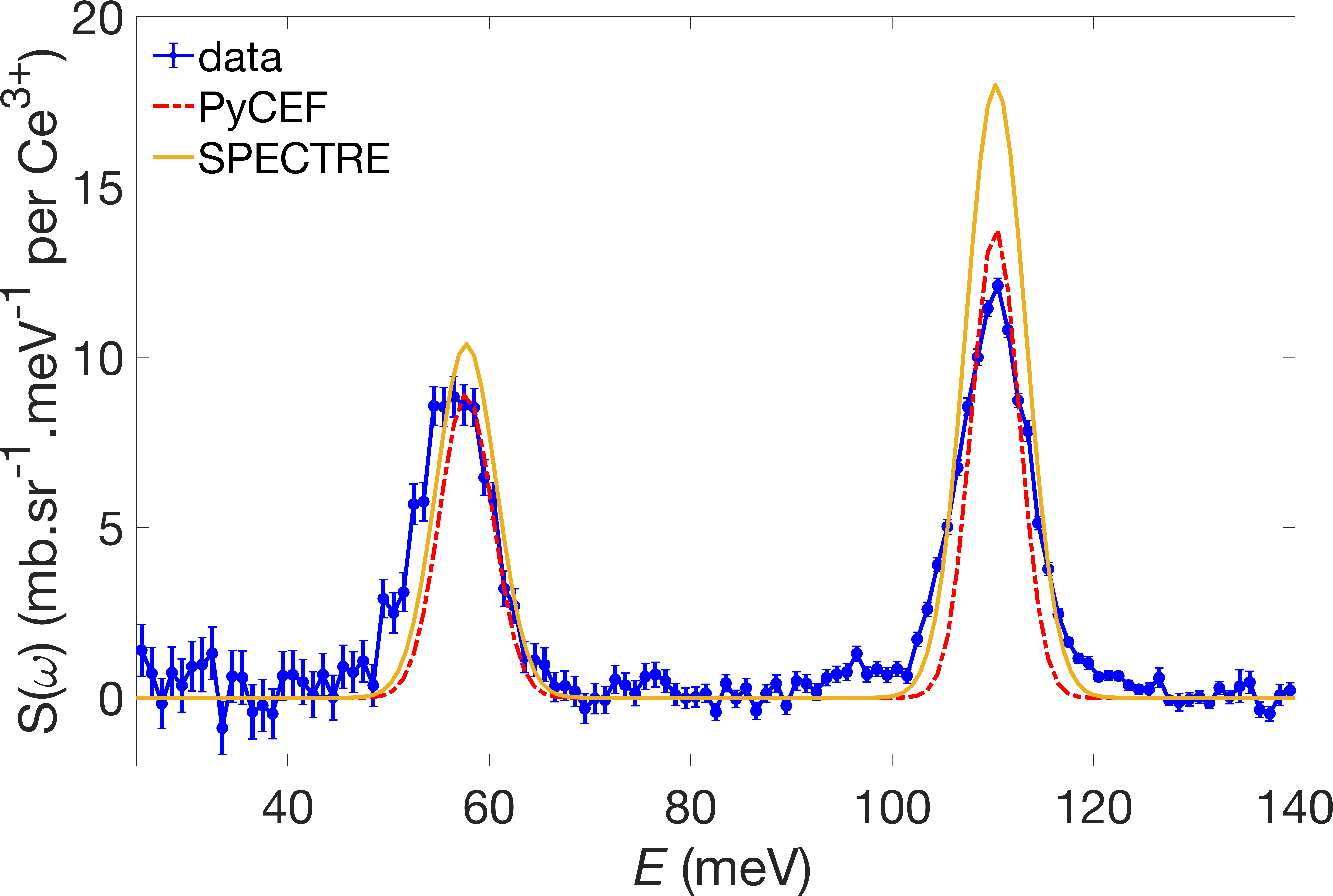}
\centering
\caption{Comparison between measured CEF excitations and calculated neutron cross section from both PyCrystalField and SPECTRE, using phenomenological energy transfer dependent peak width or constant peak width, respectively.}
\label{Fig.AppC}
\end{figure}

The flexibility of the PyCrystalField package was extremely important in our detailed analysis of the CEF data. Nevertheless, its approach is different from more classical ones based on Stevens and Wybourne formalism. SPECTRE, for instance, assumes a basis containing n electrons to be distributed within the 4\textit{f} shell. The LS coupling along with electric repulsion between electrons allows to work in the term basis, formed by the collection of $|J$,$m_{j}\rangle$, $J$ being not unique. Noteworthy, the Stevens formalism restricts to the $J$ subset of lowest energy. This results in specific correspondence between Wybourne and Stevens coefficients. PyCEF uses the L and S values of the ground state to create the basis, which gives new and specific coefficients. It is thus worth benchmarking our results with another widely used software, like SPECTRE. Using the set of CEF parameters obtained with PyCrystalField and converting it to the Wybourne formalism using internal routines, we computed the expected neutron cross section and ground-state wavefunction using SPECTRE. The wavefunction computed by SPECTRE is very similar to the one mentioned in the main text: $|\pm\rangle$ = 0.992$|^{2}F_{5/2}$,$\pm3/2\rangle$ $\mp$ 0.119$|^{2}F_{7/2}$,$\pm3/2\rangle$ $\mp$ 0.029$|^{2}F_{5/2}$,$\mp3/2\rangle$ +  0.009$|^{2}F_{7/2}$,$\mp3/2\rangle$.\\
Both expected neutron cross sections as well as the measured data are plotted in Fig.~\ref{Fig.AppC}. The small difference observed between both calculated CEF schemes potentially stems from rounding errors, different diagonalization routines and/or from the choice of formalism of each program. In addition, the simulated cross section from SPECTRE was plotted with a constant peak width whereas the PyCrystalField one uses a phenomenological energy dependence of the peak widths based on the predicted instrumental resolution and experimental data.\\

\section{Uncertainty on the crystal-electric field parameters} \label{CEF_err}
Errors on the crystal-electric field parameters in table~\ref{Table2}~were estimated using an incremental search over the variables used in the fit, following a similar method as in~\cite{scheie2021quantifying}. Due to the limited amount of observables compared to the number of CEF parameters, the fit was carried out by varying the charges used in the point charge calculations and not directly on the CEF parameters. The same applies for the incremental search which thus involves the two charges used in the point charge model (different charges for basal and apical oxygens) as well as a global scaling factor. The errors obtained on the variables were then used to determine the errors on the CEF parameters.

\begin{table*}[!h]
\begin{ruledtabular}
\small
\begin{tabular}{cccccccc}
                     &    	$|^{2}F_{5/2}$,$\mp5/2\rangle$	  & 	 $|^{2}F_{5/2}$,$\mp3/2\rangle$ 	 & 	 $|^{2}F_{5/2}$,$\mp1/2\rangle$	&	$|^{2}F_{5/2}$,$\pm1/2\rangle$	&	$|^{2}F_{5/2}$,$\pm3/2\rangle$	&	$|^{2}F_{5/2}$, $\pm5/2\rangle$   \\
			& 		$\pm$0.001				&		-0.238			&			$\mp$0.001			&			0.006			&			$\pm$0.833			&			+0.004 \\ \hline
		&	$|^{2}F_{7/2}$,$\mp5/2\rangle$  &  $|^{2}F_{7/2}$,$\mp3/2\rangle$  &  $|^{2}F_{7/2}$,$\mp1/2\rangle$    &  $|^{2}F_{7/2}$,$\pm1/2\rangle$  &  $|^{2}F_{7/2}$,$\pm3/2\rangle$ & $|^{2}F_{7/2}$,$\pm5/2\rangle$ \\ 
					&			$\mp$0.001			&			+0.155			&			$\pm$0.001			&			-0.004			&			$\mp$0.475			&			-0.002			\\
\end{tabular}
\end{ruledtabular}
\caption{\label{TableCEF2} List of contributions from the different states to the ground state wavefunction of the defective model. Note that the $|^{2}F_{7/2}$,$\pm7/2\rangle$ state is not listed here as its contribution is calculated as null.}
\end{table*}

\bibliography{CHO_Biblio}

\end{document}